# Towards Interpretable AI in Personalized Medicine: A Radiological-Biological Radiomics Dictionary Connecting Semantic Lung-RADS and imaging Radiomics Features; Dictionary LC 1.0

Ali Fathi Jouzdani[1,2], Shahram Taeb[3], Mehdi Maghsudi[1], Arman Gorji[1,2], Arman Rahmim[4,5], Mohammad R. Salmanpour[1,4,5*]

[1] Technological Virtual Collaboration (TECVICO Corp.), Vancouver, BC, Canada
[2] NAIRG, Department of Neuroscience, Hamadan University of Medical Sciences, Hamadan, Iran
[3] Department of Radiology, School of Paramedical Sciences, Guilan University of Medical Sciences, Rasht, Iran
[4] Department of Basic and Translation Research, BC Cancer Research Institute, Vancouver, BC, Canada
[5] Department of Radiology, University of British Columbia, Vancouver, BC, Canada

## ABSTRACT

**Background:** Lung cancer remains the leading cause of cancer-related mortality worldwide, with survival largely dependent on early detection. Standard-dose computed tomography (CT) screening, guided by the Lung Imaging Reporting and Data System (Lung-RADS), provides standardized criteria for nodule evaluation. However, interpretation is limited by inter-reader variability and reliance on qualitative descriptors. Radiomics offers quantitative biomarkers but faces challenges of clinical interpretability. In this work, we introduce a radiological-biological dictionary of radiomic features (RFs) that aligns quantitative metrics with Lung-RADS semantic categories, thereby bridging computational and clinical reasoning.
**Method:** We developed a clinically informed dictionary translating Lung-RADS semantic features into RFs through literature curation and expert review. As a proof of concept, imaging and clinical data from 977 patients across 12 collections from The Cancer Imaging Archive (TCIA) were analyzed. Following preprocessing and manual segmentation, 110 RFs per nodule were extracted using PyRadiomics in compliance with the Image Biomarker Standardization Initiative (IBSI). A semi-supervised learning (SSL) framework incorporating 499 labeled and 478 unlabeled cases was employed to enhance model generalizability. Seven feature selection techniques and ten interpretable classification models were evaluated. SHapley Additive exPlanations (SHAP) analysis was used to assess correspondence between feature importance and Lung-RADS descriptors.
**Results:** A clinically informed dictionary was developed through literature curation, translating ten Lung-RADS semantic features into corresponding RFs (4–24 RFs per descriptor), and was validated by eight expert reviewers. The best SSL pipeline (ANOVA + support vector machine) achieved a mean validation accuracy of 0.79±0.13. SHAP identifies key radiomic proxies corresponding to Lung-RADS descriptors (e.g., attenuation, margin irregularity, spiculation), which confirmed our dictionary approach.
**Conclusion:** The proposed dictionary provides an interpretable framework linking radiomics and Lung-RADS semantics, advancing explainable artificial intelligence for CT-based lung cancer screening.

**Keywords:** Lung-RADS; Radiomics; Explainable and Interpretable AI; Semantic mapping; Radiological-biological Radiomics dictionary; Precision oncology.

## 1. INTRODUCTION

Lung cancer is the primary cause of cancer-related deaths globally, exhibiting five-year survival rates under 20% at advanced stages of diagnosis [1]. Standard-dose computed tomography (CT) screening facilitates the early detection of nodules, thereby enhancing patient outcomes [2]. The American College of Radiology's Lung Imaging Reporting and Data System (Lung-RADS) offers standardized categories to inform follow-up and management [3, 4]. Lung-RADS mitigates inter-reader variability; however, it relies on visual assessment, which can lead to potential inconsistencies in the interpretation of borderline or heterogeneous nodules among readers [5, 6].

Quantitative radiomics provides a complementary method by extracting high-dimensional features—such as shape, intensity, and texture—from standard CT scans, thereby uncovering patterns that are not discernible to the naked eye [7]. Radiomics demonstrates utility in early detection, risk stratification, and prognosis, thereby facilitating precision screening and treatment planning [8]. Nevertheless, numerous high-performing models operate as "black boxes," which restricts clinical trust and impedes adoption [9, 10]. Explainable AI (XAI) offers post-hoc insights, such as feature importance, SHapley Additive exPlanations (SHAP) attributions, and saliency maps, to elucidate the mechanisms by which complex models generate predictions [11, 12].

Standardization remains a cornerstone of reliable radiomics research. The Image Biomarker Standardization Initiative (IBSI) and related efforts have substantially improved reproducibility in feature definitions and analytical workflows across centers and software platforms, thereby enhancing generalizability and facilitating clinical translation [13, 14]. Recent studies further support the integration of radiological semantics with biological context, connecting imaging features to morphology, tumor biology, and patient outcomes to enhance the face validity and clinical relevance of AI models [15-17]. Advancements have also been achieved through XAI tools, such as SHAP, Local Interpretable Model Agnostic Explanation (LIME), and class activation maps, which identify influential inputs; however, these tool-specific explanations alone do not provide a consistent, clinically grounded dictionary of radiomic meanings for lung nodules [18].

Despite progress in computational standardization, a fundamental challenge persists: bridging the gap between quantitative RFs and the qualitative terminology routinely used by clinicians. For instance, Lung-RADS employs intuitive descriptors such as part-solid, spiculation, and growth of ≥2 mm, whereas RFs are expressed as mathematical constructs lacking universally recognized clinical meanings. This semantic disconnect limits interpretability and hampers clinical adoption.

To address this, a series of studies were conducted in which quantitative imaging features were systematically connected to semantic descriptors derived from standardized clinical scoring systems, including PI-RADS, BI-RADS, and the World Health Organization (WHO) tumor classification. In the prostate Magnetic Resonance Imaging (MRI) study (PM1.0), first-order, texture, and shape RFs were mapped to PI-RADS-defined characteristics, including lesion intensity, heterogeneity, margins, and size thresholds [16]. In the breast MRI framework (BM1.0), IBSI-compliant features were aligned with BI-RADS lexicon terms describing lesion shape, margin, and internal enhancement patterns, and the associations were validated through SHAP-based interpretability analyses [19]. In the pathology-focused model (LCP1.0), radiomic and pathomic features were translated into WHO-derived descriptors related to architectural complexity and nuclear atypia [17]. Collectively, these efforts established a systematic linkage between standardized RADS and WHO frameworks, as well as data-driven imaging features, yielding clinically interpretable and reproducible models for multimodal disease characterization.

This study proposes an integrated framework, focusing on LC1.0, a radiological-biological dictionary that aligns Lung-RADS semantic descriptors with quantitative CT-derived RFs. The dictionary is developed through literature curation and expert review, followed by validation within machine learning (ML) frameworks for survival prediction. We utilize SHAP analyses to confirm that model-significant features correspond to anticipated Lung-RADS concepts (e.g., shape, margin, size, attenuation, growth). As mentioned in Table 1, our approach enhances transparency, reproducibility, and practical utility by linking complex model outputs to familiar clinical terminology, which is essential for establishing trustworthy AI in CT lung cancer screening.

**Table 1.** Summary of the study's contribution to existing literature.

| | |
|---|---|
| **Problem or Issue** | Lung-RADS standardizes CT nodule reporting, but interpretation remains qualitative and variable across readers; radiomics is quantitative but often hard to interpret clinically. |
| **What is Already Known** | Lung-RADS improves communication and management decisions; IBSI and PyRadiomics improve radiomics reproducibility; XAI methods (e.g., SHAP) can highlight influential features, yet do not provide a consistent semantic "translation" from radiomic features to Lung-RADS descriptors. |
| **What this Paper Adds** | Introduces LC 1.0, a clinically curated and expert-validated radiological–biological dictionary mapping 10 Lung-RADS semantic descriptors to 4–24 radiomic features per descriptor. Demonstrates, on 977 patients across 12 TCIA cohorts, that a semi-supervised pipeline (ANOVA+SVM) reaches 0.79±0.13 accuracy and that SHAP-ranked features align with Lung-RADS concepts (e.g., attenuation, margin irregularity, spiculation). |
| **Who would benefit from the new knowledge in this paper** | Radiologists, oncologists, medical physicists, and machine learning researchers are developing explainable CT screening tools; screening programs seeking standardized, auditable, and clinically grounded AI decision support. |

## 2. MATERIAL AND METHOD

### 2.1. Exploring the Relationship Between the Lung-RADS Scoring System and Descriptors

To investigate how specific imaging descriptors influence Lung-RADS categorization, the semantic features of pulmonary nodules were systematically analyzed. The descriptors included nodule shape, margin, size, attenuation, growth pattern, calcification, internal features, location, spiculation, and associated findings. Each descriptor was evaluated using standardized definitions derived from the American College of Radiology Lung-RADS version 1.1 criteria, and a comparative table was developed linking each descriptor to its corresponding Lung-RADS category and recommended clinical management action [20]. This mapping allowed for the identification of descriptor patterns most predictive of malignancy risk and follow-up intensity. Nodules characterized as round or oval with smooth margins and benign calcification were classified as Cat 1-2, indicating low risk and the need for routine annual low-dose computed tomography (LDCT). In contrast, nodules with irregular shape, spiculated margins, rapid growth, or associated lymphadenopathy were classified as Cat 4A-4X, prompting short-interval imaging or diagnostic evaluation. This structured approach facilitated a systematic assessment of the relationship between semantic imaging features and the Lung-RADS scoring framework, supporting consistent interpretation and evidence-based management in lung cancer screening.

### 2.2. Construction of the Clinically-Informed Feature Interpretation Dictionary

After establishing a mapping between radiological semantic features and Lung-RADS categories, we constructed a clinically informed feature interpretation dictionary to bridge semantic imaging descriptors with quantitative RFs. This process aimed to enhance the interpretability and clinical applicability of radiomics-based analyses in lung cancer screening. RFs extraction was performed using the PyRadiomics library [21], standardized according to the IBSI [22]. A total of 110 standardized RFs were used as the reference set, spanning multiple IBSI-defined feature classes: First-Order features (FO) (n = 19), describing voxel intensity distribution; Shape-based features (SF) (n = 17), quantifying three-dimensional lesion morphology; Gray Level Co-occurrence Matrix (GLCM) features (n = 23), capturing spatial gray-level relationships; Gray Level Size Zone Matrix (GLSZM) features (n = 16), describing zone size homogeneity; Gray Level Run Length Matrix (GLRLM) features (n = 16), quantifying the length of consecutive gray-level runs; Neighborhood Gray Tone Difference Matrix (NGTDM) features (n = 5), representing local intensity variation; and Gray Level Dependence Matrix (GLDM) features (n = 14), characterizing gray-level dependency patterns. A comprehensive list of extracted features, including definitions and abbreviations, is provided in Supplemental File 1, Supplemental Table S1.

To establish conceptual links between quantitative RFs and qualitative Lung-RADS semantic descriptors (e.g., margin, shape, attenuation, spiculation), each feature was systematically evaluated for its interpretive relevance. This process generated a clinically-informed radiomic interpretation dictionary, translating traditionally descriptive radiological terms into quantifiable imaging biomarkers. This mapping enhances both the explainability of AI-driven models and the clinical interpretability of radiomic outputs by grounding abstract mathematical features in radiological semantics [23-25]. Validation of the dictionary was performed by an expert panel consisting of three medical physicists and five clinicians (three physicians, one radiologist, and one oncologist) with extensive experience in radiomics and AI-driven image analysis. The panel first assessed the global relationships between Lung-RADS semantic attributes and the complete set of extracted RFs as candidates, then iteratively refined these associations to identify the most semantically relevant features for each descriptor.

### 2.3. Understandable Machine Learning Classification Task

The following section details the comprehensive methodology employed in this study, encompassing patient data curation from diverse multi-institutional cohorts, standardized image preprocessing and expert lesion segmentations, RFs extraction using IBSI-compliant tools, an SSL framework for survival prediction with explicit cross-validation and external validation strategies, and SHAP-based feature importance analysis aligned with Lung-RADS clinical descriptors (Figure 1).

**(i) Patient Data**. Clinical data, CT scans, and manually delineated lesion masks (as detailed in Section (iv)) were collected from 977 patients out of a total of 2,092 across 12 publicly and privately available datasets, all sourced from The Cancer Imaging Archive (TCIA). The included datasets were: LCTSC (35 patients) [26], LIDC-IDRI (279 patients) [27], Lung-Fused-CT-Pathology (6 patients) [28], LungCT-Diagnosis (48 patients) [29], NSCLC-Radiogenomics (34 patients) [30], NSCLC-Radiomics (417 patients) [30], NSCLC-Radiomics-Genomics (51 patients) [31], QIN Lung CT (35 patients) [32], RIDER Lung CT (12 patients), RIDER Pilot (4 patients) [33], SPIE-AAPM Lung CT Challenge (43 patients) [34], and TCGA-LUAD (13 patients) [29]. Demographic and clinical characteristics varied substantially across datasets. For instance, the Lung-Fused-CT-Pathology cohort comprised 83.3% males, with a mean age of 74.8 ± 10.3 years and a mean tumor size of 12.2 ± 1.9 mm. The NSCLC-Radiomics dataset comprised 72.3% males (mean age = 66.2 ± 10.3 years), with histological subtypes including squamous cell carcinoma (50.2%), large cell carcinoma (30.1%), adenocarcinoma (12.3%), and NOS/NA (7.4%). In contrast, the LungCT-Diagnosis dataset contained 59.6% of patients aged 65 years or older, with 53.2% beingmales, primarily diagnosed at advanced stages (III/IV). Of these, only NSCLC-Radiogenomics, LungCT-Diagnosis, and NSCLC-Radiomics provided survival data, with 264 patients surviving more than 2 years (Class 0) and 235 deceased within 2 years (Class 1). The remaining 478 unlabeled cases were reserved for SSL to enhance generalization through pseudo-labeling. This multi-cohort design ensured data diversity and minimized institutional bias.

**(ii) Segmentation and Preprocessing**. All CT scans passed artifact quality control and were harmonized using a standardized preprocessing pipeline. This included reconstruction to 1 mm slice thickness, resampling to 1 $mm^3$ isotropic voxels, intensity clipping to [-1000, 400] Hounsfield units (HU), min-max normalization, and 3D Gaussian smoothing ($\sigma$ = 0.5 mm). Subsequently, chest CT scans were reviewed in the standard lung window with multiplanar reconstructions to identify nodules exhibiting malignant features such as irregular shape, spiculated margins, abnormal attenuation, or rapid growth. Suspicious lesions were manually segmented on each axial slice in 3D Slicer (v5.8) by two board-certified thoracic radiologists following a standardized lung tumor contouring protocol. Pathology and multidisciplinary tumor board findings, when available, were used solely to confirm lesion identity and did not influence contour boundaries. All segmentation masks underwent independent review by a third clinical expert to ensure accuracy and inter-observer consistency. Cases with poorly defined lesion boundaries (e.g., due to effusion, fibrosis, or atelectasis) were excluded. Finalized CT volumes were then z-score normalized and masks binarized, ensuring consistent, high-quality inputs for downstream radiomics and deep learning analyses.

**(iii) RFs Extraction**. We used PyRadiomics 3.1.0, which was standardized in reference to the IBSI, with 110 standardized RFs as our reference RF set.

**(iv) Machine Learning Classification Tasks**. As a proof of concept, we implemented an ML-based classification framework for predicting survival outcomes in lung cancer. Previous studies [8, 35] have demonstrated that SSL outperforms traditional supervised learning, particularly when labeled data is limited, resulting in up to a 17% improvement in survival prediction accuracy. By leveraging unlabeled CT data, this approach offers a cost-efficient, scalable, and interpretable solution for precision oncology and clinical deployment. The labeled NSCLC-Radiomics data were used in 5-fold cross-validation, and model generalization was evaluated using two independent external cohorts, including NSCLC-Radiogenomics and LungCT-Diagnosis, to ensure cross-institutional reproducibility. Unlabeled datasets, including LCTSC, LIDC-IDRI, Lung-Fused-CT-Pathology, NSCLC-Radiomics-Genomics, QIN Lung CT, RIDER Lung CT, RIDER Pilot, SPIE-AAPM Lung CT Challenge, and TCGA-LUAD, were utilized in the SSL pseudo-labeling process to expand the training set. In the SSL procedure, the labeled NSCLC-Radiomics dataset was partitioned into five folds. In each fold, four folds were used to train a Logistic Regression Classifier (LRC), while the remaining fold served as validation.
The trained LRC was then used to generate pseudo-labels for all unlabeled samples, which were subsequently added to the training pool. A total of ten ML classification models were evaluated as final classifiers, including Logistic Regression (LR), Random Forest (RF), Extreme Gradient Boosting (XGB), Support Vector Machine (SVM), K-Nearest Neighbors (KNN), Decision Tree (DT), Naïve Bayes (NB), Adaptive Boosting (AdaBoost), Light Gradient Boosting Machine (LGBM), and Gradient Boosting Machine (GBM). Model training and validation were repeated across all five folds, and performance metrics were averaged. Feature selection was performed using seven techniques, including Analysis of Variance (ANOVA), Mutual Information (MI), Recursive Feature Elimination (RFE), L1 regularization (L1R), Variance Thresholding (VT), Extra Trees-based Importance (ETI), and Random Forest-based Importance (RFI). Hyperparameters were optimized using grid search combined with stratified 5-fold cross-validation on the labeled NSCLC-Radiomics cohort.

**(v) Feature Importance Analysis by SHAP**. Model interpretability was assessed using SHAP [36] to enhance the reproducibility and interoperability of our predictions, particularly in the context of the SSL approach. SHAP values were computed to quantify the contribution of each feature to survival prediction within the SSL framework. Survival, as a target variable, serves as a critical measure for differentiating between poor and rich prognosis, helping to classify patients based on their predicted outcomes. This approach enables a transparent understanding of how each feature affects model outputs, fostering trust and validation in the model's predictions. To assess global feature importance, we averaged SHAP rankings across the top ten ensemble models. This process identified the most significant RFs for accurate survival prediction in lung cancer.

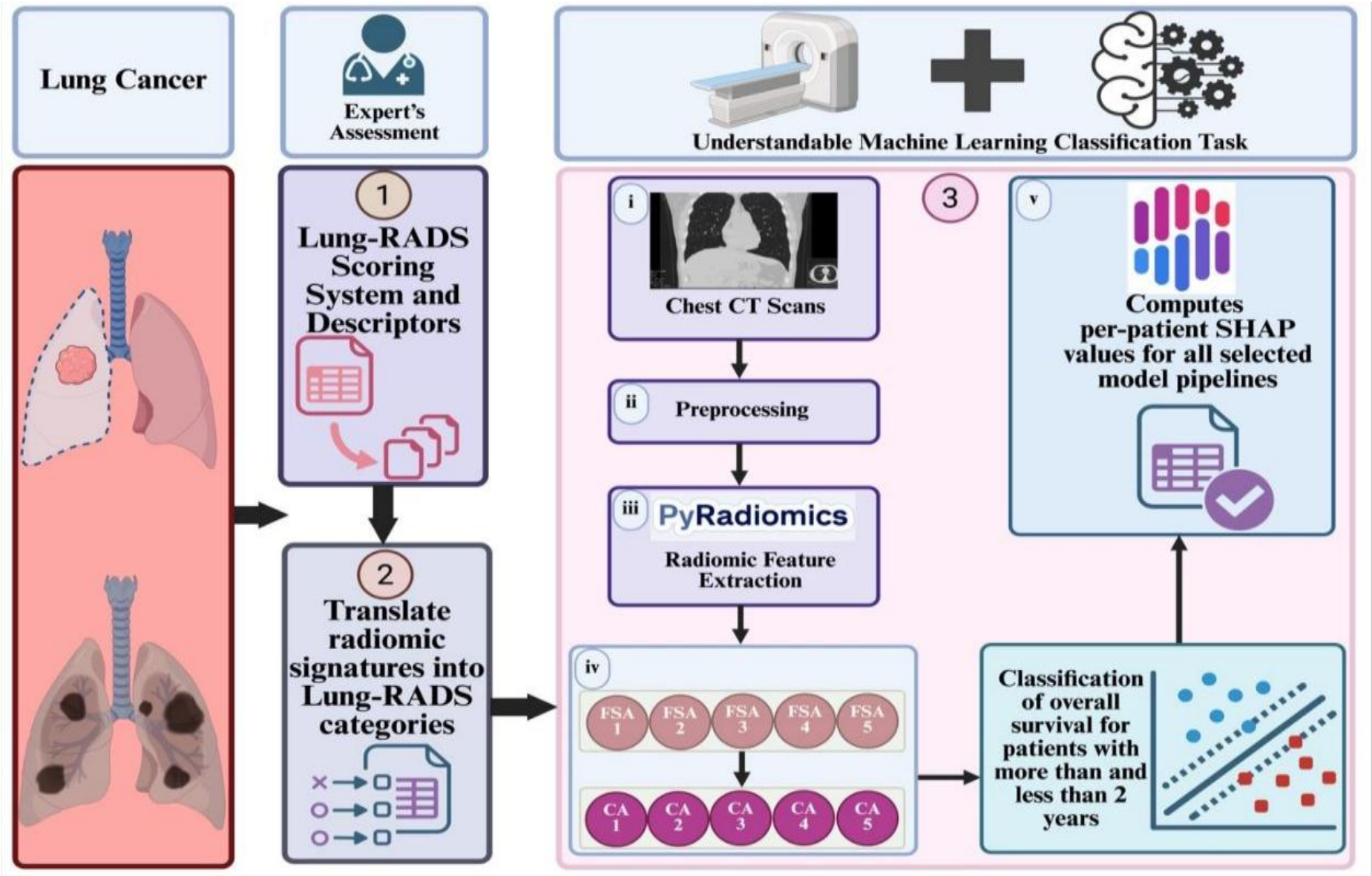

**Fig. 1.** Overview of the study workflow integrating Lung-RADS semantics, radiomics, and explainable machine learning. The workflow integrates Lung-RADS semantic descriptors with IBSI-compliant radiomic features extracted from chest CT scans to enhance interpretability and clinical relevance in survival prediction. (1) Experts evaluate Lung-RADS categories and semantic imaging features. (2) These semantic descriptors are systematically mapped to quantitative radiomic signatures by the expert panel to construct a Clinically-Informed Feature Interpretation Dictionary. (3) Chest CT scans undergo preprocessing and PyRadiomics-based feature extraction, followed by a semi-supervised learning pipeline to predict 2-year overall survival. SHAP values are then computed for the top ten models to quantify the contribution of each radiomic feature, supporting interpretability and alignment with clinical semantics. Workflow steps for the understandable machine learning classification task include: (i) Patient data acquisition; (ii) Segmentation and preprocessing; (iii) Radiomic feature extraction; (iv) Machine learning classification; (v) SHAP-based feature importance analysis. Analysis by SHAP. The resulting SHAP patterns provide complementary evidence that reinforces and confirms the expert-derived biological and radiological dictionary.

# 3. RESULTS

## 3.1. Relationship Between Lung-RADS Scoring System and Descriptors

The Lung-RADS offers a standardized framework for interpreting and managing findings in LDCT lung cancer screening. Each descriptor (nodule shape, margin, size, attenuation, growth rate, calcification pattern, internal properties, location, spiculation, and associated findings) plays a role in risk categorization and management suggestions (Table 2).

**i) Shape:** The shape of a nodule is a good sign of how likely it is to be cancerous. Lung-RADS categories 1-2 usually include round or oval nodules that are not cancerous. These nodules should be checked every year with an LDCT. On the other hand, irregular nodules indicate questionable morphology (Cat 3-4) and require a follow-up or diagnostic work-up within a short period of time.

**ii) Margin:** Benign lesions (Cat 1-2) have smooth edges, while lobulated edges make doctors suspicious that the lesion might be cancerous (Cat 3). A 6-month LDCT is needed. Spiculated or poorly defined margins are linked to the highest chance of cancer (Cat 4B/4X) and need to be checked out right away.

**iii) Size:** The size of the lesion is strongly related to how likely it is to be cancerous. Very small nodules (less than 4 mm) or small nodules (4-6 mm) usually fall under Cat 1-2 and should continue to undergo LDCT every year. Moderate-sized nodules (6-15 mm) fall into Cat 3-4A and need follow-up imaging every 3-6 months. Large nodules (>15 mm) fall into Cat 4B/4X and need a prompt diagnostic work-up.

**iv) Attenuation:** Solid nodules, especially those with bigger solid cores, are connected to a higher risk (Cat 4A-4B) and should be looked at again within three months, usually with Positron Emission Tomography (PET) imaging. Part-solid nodules (mixed attenuation) with small solid components are located in the middle of the risk scale (Cat 3-4A), and pure ground-glass nodules, especially when they are young or large, are typically classified as Cat 3, indicating that a 6-month LDCT is required.

**v) Growth Pattern:** The rate of growth is an important time-based descriptor. Stable nodules that have been there for more than a year are not dangerous (Cat 2), slow-growing lesions are not very dangerous (Cat 3), and fast-expanding nodules are very suspicious for cancer (Cat 4X) and need to be checked out right away.

**vi) Calcification Pattern:** Central, popcorn, or distributed calcifications are typical with benign conditions (Cat 1-2). However, eccentric calcification patterns are not clear (Cat 3-4A) and typically require 6 months of LDCT or more imaging.

**vii) Internal Features:** Certain internal traits also affect risk. A hamartoma (Cat 2) is indicated by fat accumulation within a nodule. Cystic nodules with thick or uneven walls are more likely to be cancerous (Cat 4A-4B) and should be checked again within three months or biopsied. Thin-walled cavitary lesions or simple cysts are usually not cancerous (Cat 2-3).

**viii) Location:** Juxtapleural or perifissural nodules are usually benign intrapulmonary lymph nodes (Cat 2), while endobronchial or central airway lesions are seen as worrisome (Cat 4A-4B) and need more testing.

**ix) Spiculation:** The level of spiculation makes risk evaluation even more precise. Absence or mild spiculation correlates with lower risk categories (Cat 1-3), but pronounced spiculation indicates a high probability of malignancy (Cat 4X) and necessitates immediate assessment.

**x) Associated Findings:** The presence of lymphadenopathy or pleural thickening markedly elevates the suspicion of cancer. If these results aren't there, the base category applies. However, if they are, the case escalates to Cat 4X, which requires full diagnostic staging.

**Table 2.** Semantic features, imaging descriptors, and their corresponding Lung-RADS implications and management recommendations. This table summarizes the relationship between key radiologic descriptors and the Lung-RADS (Lung Imaging Reporting and Data System) categories. Each semantic feature, such as nodule shape, margin, size, attenuation, growth pattern, calcification, internal features, location, spiculation, and associated findings, contributes to risk stratification and guides appropriate clinical action. Lower categories (1-2) generally indicate benign findings and annual LDCT follow-up, whereas higher categories (3-4X) denote increasing suspicion for malignancy and necessitate short-interval imaging or immediate diagnostic evaluation. Abbreviations: LDCT: Low-Dose Computed Tomography, Cat: Category, mo: month, PET: Positron Emission Tomography

| Semantic Feature | Descriptors | Lung-RADS Implication & Action |
|---|---|---|
| **Shape** | Round | Round/oval → likely benign (Cat 1-2); continue annual LDCT |
| | Oval | Round/oval → likely benign (Cat 1-2); continue annual LDCT |
| | Irregular | Irregular → suspicious (Cat 3-4); consider short-interval follow-up or work-up |
| **Margin** | Smooth | Smooth → benign (Cat 1-2); routine screening |
| | Lobulated | Lobulated → possible malignancy (Cat 3); 6-mo LDCT |
| | Spiculated | Spiculated/ill-defined → high risk (Cat 4B/4X); prompt diagnostic evaluation |
| | Ill-defined | Spiculated/ill-defined → high risk (Cat 4B/4X); prompt diagnostic evaluation |
| **Size** | Tiny (<4 mm) | Tiny/small → Cat 1-2; annual LDCT |
| | Small (4-6 mm) | Tiny/small → Cat 1-2; annual LDCT |
| | Moderate (6-15 mm) | Moderate → Cat 3-4A; 6-mo (or 3-mo for 4A) LDCT |
| | Large (>15 mm) | Large → Cat 4B/4X; immediate diagnostic work-up |
| **Attenuation** | Solid | Solid or part-solid, larger core → Cat 4A-4B; 3-mo LDCT ± PET |
| | Part-solid (mixed) | Part-solid, small core → Cat 3-4A; 3-6-mo follow-up |
| | Ground-glass | Pure ground-glass, large/new → Cat 3; 6-mo LDCT |
| **Growth** | Stable | Stable over 12 mo → benign (Cat 2); return to annual LDCT |
| | Slow-growing | Slow growth → low risk (Cat 3); 6-mo LDCT |
| | Rapidly growing | Rapid growth → high risk (Cat 4X); urgent diagnostic evaluation |
| **Calcification** | Central | Central/popcorn/diffuse → benign (Cat 1-2); routine screening |
| | Popcorn | Central/popcorn/diffuse → benign (Cat 1-2); routine screening |
| | Diffuse | Central/popcorn/diffuse → benign (Cat 1-2); routine screening |
| | Eccentric | Eccentric → indeterminate (Cat 3-4A); 6-mo LDCT or further imaging |
| **Internal Features** | Fat density | Fat → hamartoma (Cat 2); annual LDCT |
| | Cavitation | Thin-walled cavitation/simple cyst → usually benign (Cat 2-3); 6-mo LDCT |
| | Cystic change | Thick/irregular cavity walls → suspicious (Cat 4A-4B); 3-mo LDCT ± biopsy |
| **Location** | Juxtapleural/perifissural | Perifissural → benign lymph node (Cat 2); routine screening |
| | Central airway | Endobronchial/central → suspicious (Cat 4A-4B); targeted evaluation |
| **Spiculation** | Absent | Absent/minimal → lower risk (Cat 1-3); follow category schedule |
| | Minimal | Absent/minimal → lower risk (Cat 1-3); follow category schedule |
| | Marked | Marked → highest risk (Cat 4X); urgent work-up |
| **Associated Findings** | Lymphadenopathy/Pleural thickening | None → follow base category |
| | Lymphadenopathy/Pleural thickening | Present → upgrade to Cat 4X; comprehensive diagnostic staging. |

### 3.2. Clinical-Radiomic Lung-RADS Feature Mapping

The expert panel initially evaluated the broad relationships between Lung-RADS semantic features and the RFs set that covered all RFs extracted, and then iteratively refined these associations to identify the specific RFs most relevant to each semantic descriptor. Ten key Lung-RADS semantic categories, including shape, margin, size, attenuation, growth, calcification, internal characteristics, location, spiculation, and associated findings, were systematically mapped to corresponding RFs. Across categories, 4 to 24 candidate RFs were identified based on literature evidence and expert consensus [3] (Table 3; and Supplemental File 1, Supplemental Table S2).

**Location:** RFs do not explicitly represent the anatomical location of nodules. Location serves as a crucial factor in assessing malignancy risk within Lung-RADS. Juxtaplural or perifissural nodules are typically benign, whereas segmental or proximal airway nodules raise suspicion and are frequently categorized as 4A or 4B (41). This limitation underscores the need to integrate radiomic analysis with anatomical context.

**Nodule Composition (Density and Attenuation):** The brightness of a nodule on computed tomography indicates its composition. Dense, solid nodules exhibit greater brightness compared to non-solid or partially solid nodules. RFs quantify these differences using various metrics: The FO Interquartile Range quantifies the variability within the central 50 percent of voxel intensities. FO Mean Intensity represents the average gray-level value contained within the nodule. FO 10 and 90 Percentile features represent the darkest and brightest regions within the nodule. FO Energy and Total Energy quantify voxel intensity magnitudes, where elevated values signify bright, solid nodules. FO Maximum Intensity, Median Intensity, and Minimum Intensity define the spectrum and equilibrium of gray levels. Texture-based features offer additional understanding of heterogeneity. The metrics include the GLCM Sum Average and Joint Average, which assess brightness and gray-level pair distributions; GLDM Low Gray-Level

Emphasis and High Gray-Level Emphasis, which evaluate the prevalence of dark versus bright voxels; and the GLSZM High Gray-Level Zone Emphasis and Low Gray-Level Zone Emphasis, GLRLM High Gray-Level Run Emphasis and Low Gray-Level Run Emphasis, which characterize the prevalence of bright and dark runs or zones. These features collectively offer quantitative proxies for the Lung-RADS classifications of solid, part-solid, and non-solid nodules (Table 3, Attenuation rows; and Supplemental File 1, Supplemental Table S2).

**Size (Diameter and Volume):** Size constitutes a fundamental criterion within Lung-RADS. Radiomic SFs quantitatively represent diameters and volumes. The SF Maximum Three-Dimensional Diameter quantifies the most significant Euclidean distance between points on the surface of a nodule. The SF Maximum Two-Dimensional Diameters represent the greatest extent across the axial, coronal, and sagittal planes. The SF Maximum Axis Length indicates the length of the principal ellipsoid axis that encompasses the nodule. SF Mesh Volume and Voxel Volume provide direct volumetric estimates. The features correspond closely with Lung-RADS thresholds, including the ≥ 15 mm cut-off for higher-risk categories (Table 3, Sizes rows; and Supplemental File 1, Supplemental Table S2)

**Growth Behavior:** No individual RFs explicitly represent temporal growth; however, numerous features exhibit variations as nodules increase in size or exhibit greater heterogeneity. Delta-radiomics monitors longitudinal alterations in characteristics such as maximum three-dimensional diameter, voxel volume, and textural heterogeneity, offering quantitative evidence of growth. This is consistent with Lung-RADS criteria, as documented interval growth significantly heightens the suspicion of malignancy. Rapidly enlarging nodules showed significant increases in SF Delta Voxel Volume over time, correlating with aggressive growth patterns (Lung-RADS 4X). Volume change served as a sensitive marker for progression and risk of malignancy (Supplemental File 1, Supplemental Table S2).

**Morphology and Shape:** Round and oval nodules exhibited high SF Sphericity and moderate SF Major Axis Length values, indicating compact, ellipsoid forms typical of benign lesions (Lung-RADS 1-2). Conversely, irregular nodules demonstrated elevated spherical disproportion, consistent with asymmetric or distorted morphologies frequently observed in suspicious categories (Lung-RADS 3-4). Smooth margins were characterized by high SF Compactness 2, confirming well-defined, rounded contours associated with low-risk nodules (Lung-RADS 1-2). Lobulated margins showed reduced SF Elongation, indicating undulating or segmented borders often seen in indeterminate nodules (Lung-RADS 3). Markedly spiculated margins presented an elevated SF Surface-to-Area Volume Ratio, reflecting complex, irregular outlines and strong association with malignancy (Lung-RADS 4B/4X) (Table 3, Shape rows; and Supplemental File 1, Supplemental Table S2).

**Internal Features:** Internal findings such as calcification, fat, and cavitation significantly affect Lung-RADS scoring. Calcified nodules manifest as highly luminous areas exhibiting significant contrast relative to adjacent tissue. RFs, including FO Maximum Intensity, FO Energy, and NGTDM Contrast, effectively capture these properties. Calcified nodules typically demonstrate high intensity and pronounced local contrast, whereas non-calcified solid nodules, despite being bright, may lack such distinct contrast (Table 3, Internal Features; and Supplemental File 1, Supplemental Table S2).

**Associated Findings (Category 4X):** Lung-RADS Category 4X encompasses complex associated findings, including pleural thickening and lymphadenopathy, which heighten the suspicion of malignancy. Numerous higher-order RFs closely correspond to these descriptors. Lower values of GLCM Informational Measure of Correlation indicate disorganized and invasive tissue characteristics. GLDM Dependence Non-Uniformity Normalized: elevated values signify non-uniform dependence patterns, aligning with secondary findings such as pleural thickening or nodal involvement. GLCM Inverse Difference Normalized: lower values indicate decreased homogeneity, associated with fuzzy, poorly defined margins necessitating prompt assessment. NGTDM Contrast: elevated values signify pronounced local intensity variations between voxels and their neighbors, effectively capturing highly irregular and ambiguous boundaries. These features serve as quantitative proxies for the suspicious findings that support Category 4X classification in Lung-RADS (Table 3, Associated Findings rows; and Supplemental File 1, Supplemental Table S2).

**Table 3.** Mapping of Lung-RADS semantic descriptors to corresponding RFs. Each feature quantifies morphological, textural, and intensity-based characteristics underlying nodule categorization and malignancy risk.

| Lung-RADS | Semantic Descriptor | Radiomic Feature | Meaning |
|---|---|---|---|
| **Category 1-2** | Shape: Round / Oval | Sphericity (SF) | Closer to perfect sphere |
| | | Compactness1 (SF) | Compactness relative to a sphere |
| | | Compactness2 (SF) | Alternate compactness metric |
| | | Surface-Area to Volume Ratio (SF) | SA/V; lower → more compact |
| | | Maximum 3D Diameter (SF) | Largest surface distance |
| | Margin: Smooth | Inverse Difference Moment (GLCM) | Local homogeneity |
| | | Inverse Difference Normalized (GLCM) | Normalized homogeneity |
| | | Cluster Shade (GLCM) | Unevenness in tissue intensity clustering |
| | Size: Tiny (<4 mm) / Small (4-6 mm) | Voxel Volume (SF) | ROI volume |
| | | Mesh Volume (SF) | Mesh-derived volume |
| | | Maximum 3D Diameter (SF) | Largest span in any direction |
| | Calcification: Central / Popcorn / Diffuse | Cluster Prominence (GLCM) | Sharp intensity differences in clustered regions |
| | | Cluster Tendency (GLCM) | clustering Strength of similar gray levels |
| | | Homogeneity (GLCM) | Local uniformity |
| | Internal: Fat density (hamartoma) | Low Gray-Level Zone Emphasis (GLSZM) | Emphasis on low-intensity zones |
| | | Small Area Low Gray-Level Emphasis (GLSZM) | Small & low-intensity zones |
| | Internal: Thin-walled cavity / Simple cyst | Variance (FO) | Spread of intensities |
| | | Range (FO) | Max-min intensity separation |
| | Spiculation: Absent / Minimal | Inverse Difference Moment (GLCM) | Local homogeneity |
| | | Short-Run Emphasis (GLRLM) | Emphasis on short, uniform runs |
| **Category 3** | Shape: Irregular | Spherical Disproportion (SF) | ↑ SpD ↔ more irregular |
| | | Flatness (SF) | ↓ Flatness ↔ irregular |
| | | Elongation (SF) | ↓ Elongation ↔ irregular |
| | Margin: Lobulated | Cluster Tendency (GLCM) | ↓ CT ↔ lobulated |
| | | Contrast (GLCM) | mid-range ↔ lobulation |
| | Size: Moderate (6-15 mm) | Max 3D Diameter (SF) | 6-15 mm ↔ moderate |
| | | Mesh Volume (SF) | mid-range volume ↔ moderate |
| | Attenuation: Part-solid | 10th Percentile (FO) | ↓ 10P ↔ low-density core |
| | | Mean Absolute Deviation (FO) | ↑ MAD ↔ mixed intensities |
| | Attenuation: Ground-glass | Gray-Level Variance (GLSZM) | ↑ GLV ↔ heterogeneous GGO |
| | | Entropy (FO) | ↑ En ↔ randomness |
| | Calcification: Eccentric | Inverse Difference (GLCM) | ↓ ID ↔ eccentric calcification |
| | | Long Run Emphasis (GLRLM) | ↑ LRE ↔ long, uneven streaks |
| | Internal: Cavitation (thin-walled) | Variance (FO) | low variance ↔ uniform cavity |
| | | Range (FO) | low range ↔ narrow intensity |
| | Spiculation: Absent/Minimal | Inverse Difference Moment (GLCM) | ↑ IDM ↔ smooth margin |
| | | Short-Run Emphasis (GLRLM) | ↓ SRE ↔ few short spikes |
| **Category 4A** | Size: Moderate (6-15 mm) | Max 3D Diameter (SF) | 6-15 mm ↔ moderate |
| | | Mesh Volume (SF) | mid-range volume ↔ moderate |
| | Attenuation: Solid / Large core | Mean Intensity (FO) | ↑ MI ↔ solid density |
| | | 90th Percentile (FO) | ↑ 90P ↔ bright core |
| | Attenuation: Part-solid (large core) | Mean Absolute Deviation (FO) | ↑ MAD ↔ heterogeneous core |
| | | Uniformity (FO) | ↓ Un ↔ heterogeneous core |
| | Internal: Thick/irregular cavity walls | Contrast (GLCM) | ↑ Contrast ↔ wall irregularity |
| | | Run Entropy (GLRLM) | ↑ REn ↔ Intensity patterns vary more across tissue |
| | Spiculation: Marked | Short-Run Emphasis (GLRLM) | ↑ SRE ↔ many short spikes |
| | | Contrast (GLCM) | ↑ Contrast ↔ sharp spikes |
| | Associated: Present | Joint Entropy (GLCM) | ↑ JEn ↔ complex secondary signs |
| | | Zone Percentage (GLSZM) | ↑ ZP ↔ fine-texture associations |
| **Category 4B** | Size: Large (>15 mm) | Voxel Volume (SF) | ↑ VV ↔ large volume |
| | | Mesh Volume (SF) | ↑ MV ↔ large volume |
| | | Surface-Area/Volume (SF) | ↑ SAVR ↔ irregular large shape |
| | Attenuation: Solid / Large core | Standard Deviation (FO) | ↑ SD ↔ dense solid region |
| | Margin: Spiculated / Ill-defined | Short-Run Emphasis (GLRLM) | ↑ SRE ↔ dense spiculation |
| | | Contrast (GLCM) | ↑ Contrast ↔ sharp, spiculated margins |
| | | High Gray-Level Zone Emphasis (GLSZM) | ↑ HGLZE ↔ bright spike zones |
| **Category 4X** | Growth: Rapidly growing | Percent Δ Volume | large % Δ ↔ aggressive growth |
| | | Area Δ (SF) | ↑ ΔArea ↔ rapid expansion |
| | Spiculation: Marked | Short-Run Low Gray-Level Emphasis (GLRLM) | ↑ SRLGLE ↔ dense, low-intensity spikes |
| | | Zone Variance (GLSZM) | ↑ ZV ↔ highly variable zone sizes |
| | Associated Findings: Present | Informational Measure of Correlation 1 (GLCM) | ↓ IMC1 ↔ complex invasive relationships |
| | | Dependence Non-Uniformity Normalized (GLDM) | ↑ DNN ↔ non-uniform dependencies |
| | Margin: Ill-defined | Inverse Difference Normalized (GLCM) | ↓ IDN ↔ ill-defined margin |
| | | NGTDM Contrast (NGTDM) | ↑ Con ↔ high spatial change in intensity |

Additionally, more details about the proposed relationships between semantic features and radiomics are provided in Supplemental File 1, Supplemental Table S3.

## 3.3. Understandable Classification Task

### 3.3.1. Classification

We assessed the applicability of RFs for prognostic stratification by testing various SSL pipelines that integrate different feature selection strategies, pseudo-classifiers, and final models. Performance was evaluated by calculating the mean validation accuracy over five folds, along with the associated standard deviations. The optimal pipeline was achieved through the combination of ANOVA feature selection, LRC as the pseudo-labeling classifier, and SVM as the final model, resulting in a mean validation accuracy of 0.79 ± 0.13. This model consistently surpassed other combinations, indicating that ANOVA-based feature selection is effective in minimizing redundancy in RFs while preserving discriminative power. LRC, functioning as both a pseudo-classifier and final classifier, attained a performance metric of 0.78 ± 0.15.

In contrast, the RF, serving as the final model, achieved a performance metric of 0.78 ± 0.13. Despite a marginal decrease in accuracy, these models demonstrated competitive and stable performance. Tree-based feature importance methods, including RFI and ETI, produced moderately strong results, achieving accuracies between 0.65 and 0.75, though they exhibited increased variability across folds. (Figure 2). Additional performance metrics and model hyperparameters are provided in Supplemental File 2.

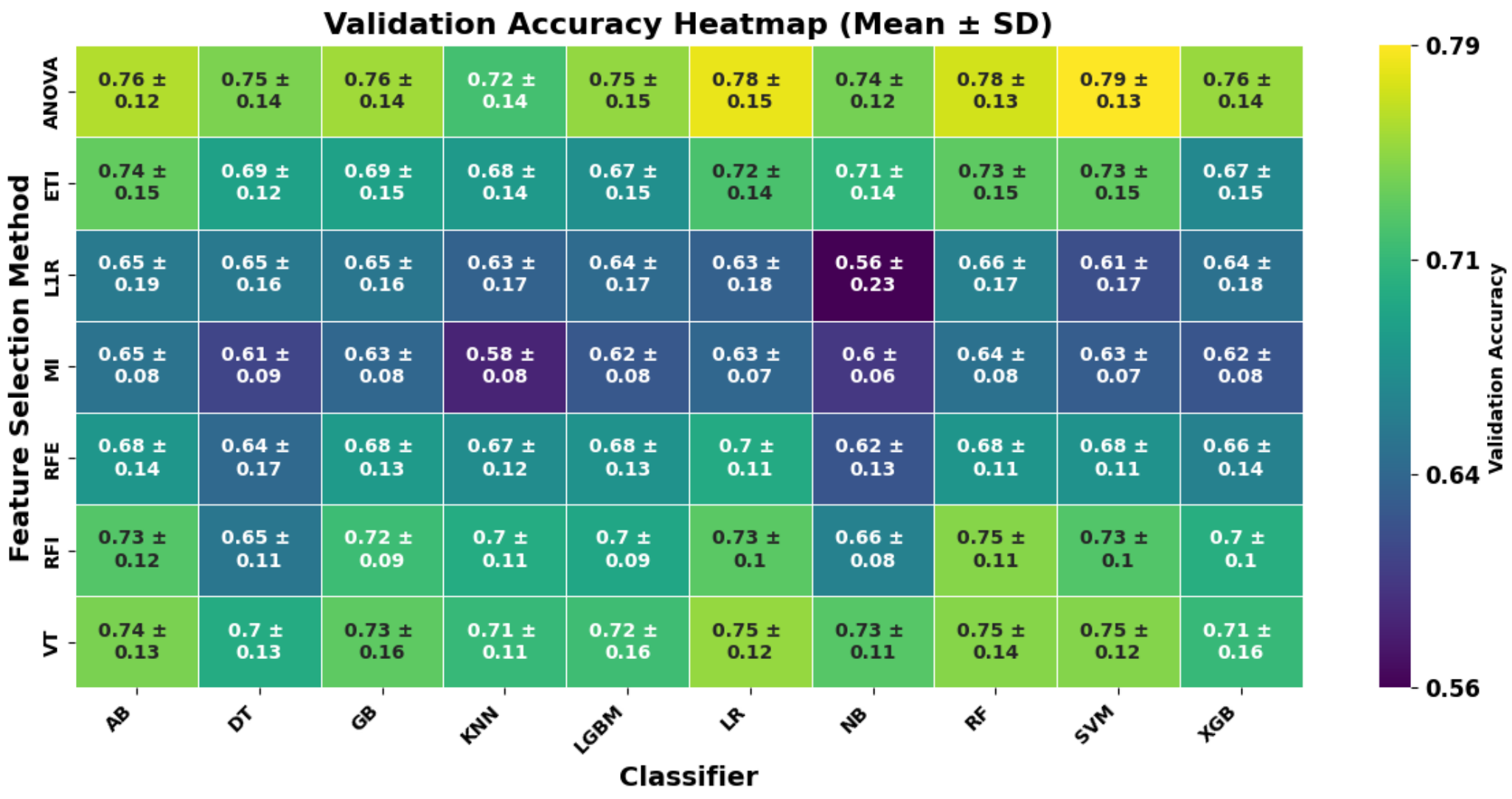

**Fig. 2.** Heatmap of validation performance across feature selection methods and classifiers. LR: Logistic Regression, RF: Random Forest, XGB: Extreme Gradient Boosting, SVM: Support Vector Machine, KNN: K-Nearest Neighbors, DT: Decision Tree, NB: Naïve Bayes, AB: Adaptive Boosting, LGBM: Light Gradient Boosting Machine, GBM: Gradient Boosting Machine, ANOVA: Analysis of Variance, MI: Mutual Information, RFE: Recursive Feature Elimination, L1R: L1 regularization, VT: Variance Thresholding, ETI: Extra Trees-based Importance, RFI: Random Forest-based Importance

### 3.3.2. SHAP-Based Validation of the Lung-RADS-Radiomics Feature Dictionary

As shown in Figure 3, the top ten SHAP-ranked features indicate that poor-prognosis classification is driven primarily by texture-based heterogeneity rather than gross morphology. The strongest positive contributor was GLCM Joint Entropy, with higher values increasing the model-predicted probability of poor prognosis. According to our Lung-RADS-radiomics feature dictionary, elevated GLCM Joint Entropy reflects complex secondary signs mapped to Lung-RADS category 4A, which is associated with poorer outcomes. The FO 90th percentile intensity also emerged as an important dictionary-mapped feature; however, higher values were associated with a reduced probability of poor prognosis. This directionality is consistent with our dictionary translation, in which this feature is mapped exclusively to Lung-RADS category 4B. Furthermore, higher values of GLRLM Run Entropy were associated with poor prognosis, aligning with its dictionary mapping to internal heterogeneity and thick or irregular cavity walls in Lung-RADS category 4A. Additional contributors, including GLCM Difference Entropy and Joint Average, are not currently represented in the dictionary. A comprehensive list of SHAP feature importances is provided in Supplemental File 3.

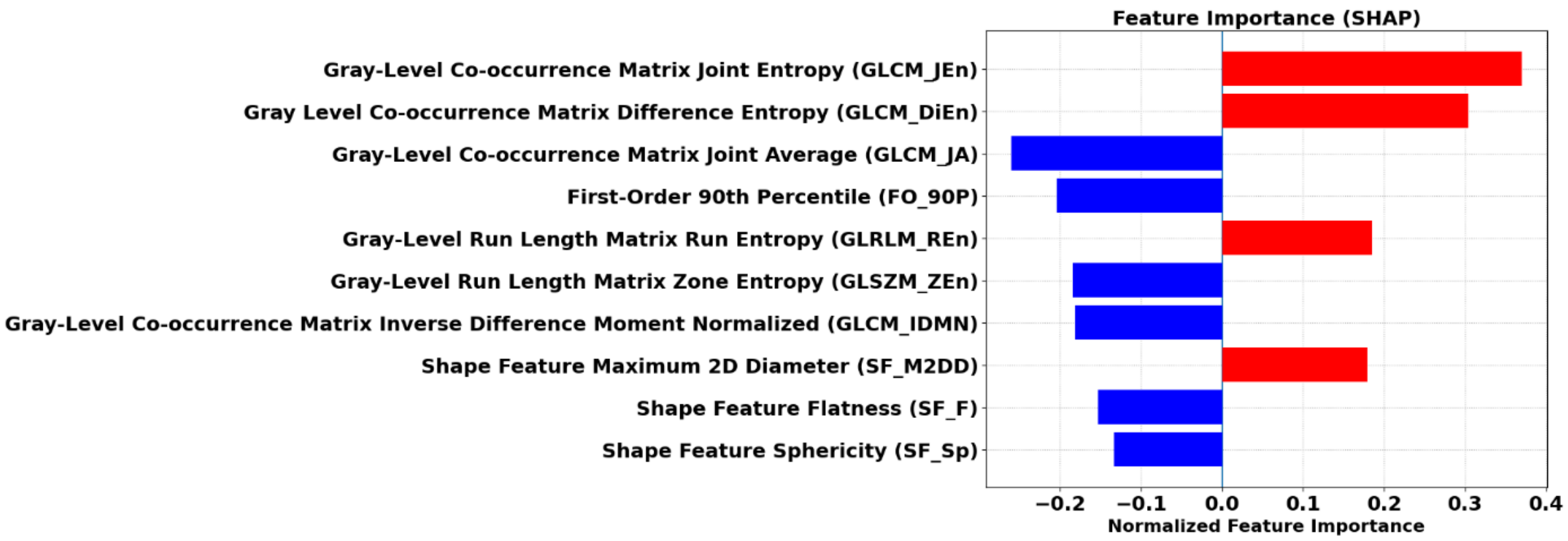

**Fig. 3.** Normalized SHAP feature importance for radiomic predictors of survival in lung cancer screening. Positive (red) values increase the prediction of Class 1 (poor prognosis/short-term survival), whereas negative (blue) values decrease it.

## 4. DISCUSSION

The adoption of artificial intelligence in lung cancer screening is often hindered by the opaque nature of many models. Standardized clinical frameworks, such as Lung-RADS, offer a common vocabulary for communicating risk from category 0 to category 4X, taking into account size, attenuation, growth, and morphology. This makes them a perfect platform for explainable algorithms [37, 38]. By matching RFs to Lung-RADS semantic features, we turn vague assessments of intensity, shape, and texture into objective stand-ins for categories that radiologists already use [39-41]. This makes things clearer and easier to act on at the point of care. The radiological and biological dictionary defines ten Lung-RADS descriptor families: shape, margin, size, attenuation, growth, calcification, cavitation, location, spiculation, and associated findings, utilizing quantitative radiomic proxies [15, 22, 42]. SHAP analyses confirm that the features influencing predictions align with the relevant clinical concepts [11, 43]. In practical applications, FO_E and total energy, along with high percentile intensity, correlate with solid density in categories three through 4B. Meanwhile, the gray level size zone matrix, large area high gray level emphasis, and zone variance effectively represent dense cores, irregular margins, and heterogeneity characteristic of categories 4B and 4X [23, 44]. Conversely, a large area of low gray level emphasis corresponds to benign, ground-glass-dominant appearances in categories one and two [3]. This coupling advances from ad hoc visualizations to auditable explanations anchored in a dictionary framework [23, 30].

As an illustrative application of the proposed dictionary, the combination of ANOVA-based feature selection with a support vector machine achieved the highest classification performance, with a mean validation accuracy of 0.79 ± 0.13. Importantly, this pipeline enabled interpretable mapping of model-selected RFs to Lung-RADS semantic descriptors. SHAP analysis indicated that poor-prognosis classification was primarily driven by texture-based heterogeneity rather than gross morphology, with GLCM Joint Entropy emerging as the strongest positive contributor, corresponding to complex secondary signs mapped to Lung-RADS category 4A. In contrast, higher values of the FO 90th percentile intensity were associated with a reduced probability of poor prognosis, consistent with its exclusive mapping to Lung-RADS category 4B and highlighting the framework's ability to capture distinct radiological phenotypes. Similarly, elevated GLRLM Run Entropy aligned with internal heterogeneity and thick or irregular cavity walls in Lung-RADS category 4A, reinforcing the clinical coherence of the dictionary. Additional influential features, including GLCM Difference Entropy and Joint Average, were not currently represented in the dictionary, suggesting directions for future refinement and expansion. Together, these radiomic-semantic relationships provide clinically meaningful insights into tumor aggressiveness and support risk-stratification frameworks used in routine patient management. This enabled the model's RFs to be interpreted directly within the Lung-RADS semantic framework, clarifying why specific features contributed to predictions of favorable or poor 2-year survival.

The RFs highlighted by our models are robustly substantiated. Previous studies [45-47] have indicated that texture correlation is associated with lower values in malignancy, sphericity shows lower values in malignant nodules, surface to volume ratio exhibits higher values for irregular margins, maximum two-dimensional diameter is relevant, and small area emphasis pertains to fine texture. Additional studies [48-50] identified discriminative shape features, including area, perimeter, and major and minor axes, alongside traditional texture families derived from gray level co-occurrence, dependence, and run length matrices. Further studies [45, 51] identified four consistently valuable features: sphericity, gray level co-occurrence maximum probability (including Laplacian of Gaussian or wavelet-scaled variants), first-order ninetieth percentile, and surface-to-volume ratio. The results from our dictionary and SHAP analysis support these findings and offer a Lung-RADS-based rationale for their clinical significance.

Mapping RFs to Lung-RADS facilitates specific applications. Examples encompass the quantification of margin irregularity and spiculation to facilitate category 4B or 4X upgrades, the standardization of attenuation measurements for part-solid versus solid adjudication, the identification of rapid change through delta radiomics as an additional growth criterion, and the production of structured Lung-RADS-aligned reports that incorporate SHAP-ranked features and confidence statements [52]. This synergy enhances risk stratification by differentiating between a stable category two nodule and a genuinely suspicious category four nodule, thereby increasing clinician confidence and facilitating integration into standard low-dose CT workflows [53, 54].

SHAP identified additional signals, including gray level co-occurrence skewness and directional or joint entropy, which are not explicitly mentioned in Lung-RADS but may effectively represent complex multiscale heterogeneity [23, 41]. Consequently, we implement a conservative policy. Clinical mappings are prioritized over suggestions based solely on SHAP. Features identified as highly important SHAP are regarded as hypotheses for prospective validation and potential candidates for future dictionary expansion. The clinical dictionary developed in this study holds significant potential for broadening the scope of outcome predictions in clinical practice. Although the dictionary's RFs were defined based on Lung-RADS, its methodology is inherently expandable to predict a variety of other clinical outcomes beyond survival, such as treatment response, disease recurrence, progression-free survival, and recurrence-free survival in lung cancer or other conditions. Lung-RADS, which categorizes lung nodules based on their radiological characteristics, is closely linked to survival, progression-free survival, recurrence-free survival, and TNM staging, as it helps assess the malignancy risk of detected nodules and stratifies patients accordingly [55]. Given its ability to integrate imaging and molecular data, the dictionary can adapt to diverse clinical settings, providing a more comprehensive understanding of disease dynamics. In clinical practice, this dictionary could assist in personalized treatment planning, enabling physicians to make more informed decisions based on predictions of various outcomes. Furthermore, its flexibility makes it applicable to other diseases, broadening its potential use in precision medicine and offering a scalable, data-driven tool for clinical decision-making.

There are several limitations to our work. The LC 1.0 was developed using a limited number of physicians and lacks validation across institutions or populations. Broader validation with more clinicians is needed to refine the clinical dictionary, and larger, more diverse datasets will improve generalizability. Although the RFs were defined based on Lung-RADS, they can be further validated using other scoring systems, such as the Tumor, Node, Metastasis (TNM) staging system, to enhance their robustness and applicability. Exploring alternative model architectures and feature selection strategies may enhance predictive performance. Integrating RFs with other modalities, such as transcriptomics or digital pathology, could yield more robust and biologically meaningful models. Additionally, prior studies emphasize the importance of assessing RF stability across different segmentations and imaging conditions, as variability can affect reliability. Future work aims to expand the framework using significantly more RFs extracted through PySERA [56], which will enable more standardized, explainable, and reproducible pipelines. Prospective validation and real-time clinical feedback will be key for successful clinical integration.

The dictionary was curated by experts and validated retrospectively. Multicenter prospective studies are necessary to evaluate transportability and clinical significance. SHAP interpretations are contingent upon the model used and will achieve stability with the inclusion of larger, more diverse cohorts and ongoing harmonization efforts. Future work will involve the release of reference software for automatic Lung-RADS aligned quantitative reporting, the integration of delta radiomics to explicitly model growth, the calibration of thresholds for decision support, and the execution of reader-in-the-loop studies to quantify improvements in inter-reader agreement and workflow efficiency.

# 5. CONCLUSION

By aligning quantitative RFs with Lung-RADS semantic descriptors, the proposed clinical dictionary bridges mathematical image biomarkers with radiologist-driven concepts such as size, attenuation, margin irregularity, and heterogeneity, and the present study serves as a concrete example of how this dictionary can be used to interpret imaging features clinically. Within this framework, the best-performing pipeline (ANOVA + SVM) achieved strong generalization across multi-institutional datasets, reaching a mean validation accuracy of $0.79 \pm 0.13$. SHAP analyses demonstrated that the most influential features mapped consistently to clinically meaningful Lung-RADS descriptors, reinforcing the interpretability of model predictions. Together, these results show how dictionary-guided interpretability can enhance clinical trust and usability. In clinical practice, this Lung-RADS-aligned dictionary can support structured, explainable decision-making by translating quantitative imaging features into standardized radiological descriptors that enhance risk stratification, reporting consistency, and confidence in lung cancer screening and follow-up.

**DATA AND CODE AVAILABILITY.** All codes and tables are publicly shared at:

https://github.com/MohammadRSalmanpour/A-Radiological-Biological-Radiomics-Dictionary-for-Lung-Caancer-Dictionary-LC-1.0

**ACKNOWLEDGMENTS.** This study was supported by the Natural Sciences and Engineering Research Council of Canada (NSERC) Discovery Horizons Grant DH-2025-00119. This study was also supported by the Virtual Collaboration Group (VirCollab.com) and the Technological Virtual Collaboration (TECVICO CORP.) based in Vancouver, Canada.

**CONFLICT OF INTEREST.** Authors Drs. Ali Fathi Jouzdani, Mehdi Maghsudi, Arman Gorji, and Mohammad R. Salmanpour were employed by the company Technological Virtual Collaboration (TECVICO Corp.). The other co-authors declare no relevant conflicts of interest or disclosures.